# Disorder Induced Ferromagnetism in CaRuO$_3$

T. He and R.J. Cava

*Department of Chemistry and Princeton Materials Institute, Princeton University*

*Princeton, New Jersey 08540, U.S.A.*

## Abstract

The magnetic ground state of perovskite structure CaRuO$_3$ has been enigmatic for decades. Here we show that paramagnetic CaRuO$_3$ can be made ferromagnetic by very small amounts of partial substitution of Ru by Ti. Magnetic hysteresis loops are observed at 5 K for as little as 2% Ti substitution. Ti is non-magnetic and isovalent with Ru, indicating that the primary effect of the substitution is the disruption of the magnetic ground state of CaRuO$_3$ through disorder. The data suggest that CaRuO$_3$ is poised at a critical point between ferromagnetic and paramagnetic ground states.

## I. INTRODUCTION

The ruthenate perovskites and their layered variants provide an important testing ground for current ideas in the magnetism of complex systems, and in particular, the possible relationships between magnetism and superconductivity. [1-9] The changes from ferromagnetism to superconductivity on going from three-dimensional SrRuO$_3$ to two-dimensional Sr$_2$RuO$_4$, [1-4] and from ferromagnetism to paramagnetism on going from SrRuO$_3$ to CaRuO$_3$ [10-13] are important cornerstones for defining the fundamental character of the whole family. CaRuO$_3$, cited in past discussions as a paramagnetic material with antiferromagnetic spin interactions, establishing a baseline for "normal" magnetic behavior in this system, is in actuality quite problematic. Recently, a new context for understanding CaRuO$_3$ has been proposed, in which CaRuO$_3$ is considered to be a nearly ferromagnetic metal, rather than a classical Curie-Weiss antiferromagnet. [14-17] Here we show that small amounts of disorder introduced into CaRuO$_3$ by partial substitution of Ru by non-magnetic, isovalent Ti (Ti$^{4+}$ has electron configuration $3d^0$) induce clear ferromagnetic behavior at low temperatures: magnetic hysteresis loops are observed at 5 K for as little as 2% Ti substitution. CaRuO$_3$ itself appears to show magnetic behavior which is part of a continuous trend towards increasing ferromagnetism induced by increasing disorder, and thus appears to be poised at a critical point between ferromagnetism and paramagnetism at low temperatures.

## II. EXPERIMENTAL

Samples of CaRu$_{1-x}$Ti$_x$O$_3$ (0≤x≤1) were prepared by conventional solid state reaction using CaCO$_3$ (99.8% Mallinckrodt), dried RuO$_2$ (99.95% Cerac) and TiO$_2$ (99.9+% Alfa). The starting materials were mixed in stoichiometric proportion, and heated in dense Al$_2$O$_3$ crucibles at 1000°C in air for three days with intermediate grindings. About 5% wt. KCl was added to the samples as a reaction rate enhancer. The powders were then pressed into pellets and heated in dense Al$_2$O$_3$ boats at 1250°C in air for two days with intermediate grindings. Samples were investigated by powder X-ray diffraction with Cu Kα radiation at room temperature; a small amount of fine tungsten powder was mixed with each ground sample as an internal standard. The cell parameters were refined by least squares fitting of powder X-ray diffraction reflection positions.

Electrical resistivity was measured in the range of 5-300 K using a standard four-lead AC technique on polycrystalline sample bars about 1x1x5 mm$^3$ in size. The magnetic properties were studied in a Quantum Design PPMS system. Magnetic susceptibility data were collected in the temperature range 5-300 K in a DC field of 1 Tesla on field cooling. The magnetic hysteresis loops were taken in 5 or 10 K steps between 5 K and 55 K for each sample, in a magnetic field ranging from −2 to 2 Tesla.

## III. RESULTS AND DISCUSSION

The X-ray patterns show that all samples of CaRu$_{1-x}$Ti$_x$O$_3$ (0≤x≤1) have the GdFeO$_3$-type orthorhombic perovskite structure with no secondary phases detected. The refined cell parameters are shown in the main panel of Fig. 1: $a$ and $b$ decrease, while $c$ increases smoothly when Ru$^{4+}$(0.62Å) is substituted by smaller Ti$^{4+}$(0.605Å). The inset to Fig. 1 gives the temperature dependence of the resistivity for selected compositions of polycrystalline samples of CaRu$_{1-x}$Ti$_x$O$_3$. CaRuO$_3$ is metallic, while the resistivity of CaRu$_{1-x}$Ti$_x$O$_3$ goes up systematically on Ti-doping. The grain boundary resistance has a significant effect on the observed resistivity, so no quantitative conclusions can be drawn from the data. However the observed transition from metallic CaRuO$_3$ to insulating CaTiO$_3$ is as expected, as Ti$^{4+}$ has no $d$ electrons and Ti-O hybridization is poor.

The temperature dependence of inverse magnetic susceptibility measured at 1 T on field cooling is shown in Fig. 2. The inverse magnetic susceptibility decreases systematically when x goes from 0 to 0.4, with slightly increasing slopes in the high temperature region. When x is greater than 0.4, however, $1/\chi$ begins to increase as shown in the inset of Fig. 2. ($CaTiO_3$ is diamagnetic and is excluded from the figure.) The high temperature data can be well described by fitting to the Curie-Weiss law. All samples, however, including $CaRuO_3$ itself, show a systematic S-shaped downward trend below 50 K. Similar behavior has been observed in Sr doped $CaRuO_3$, and was attributed to the occurrence of Sr clusters in the solid solution, leading to short-range ordered ferromagnetic regions of $SrRuO_3$. [18] There are no Sr ions present in the current system. The same type of short-range ferromagnetic behavior has therefore been introduced in $CaRuO_3$ by Ti doping.

The Curie-Weiss temperature, $\theta_{CW}$, derived by fitting the 150-300 K data in Fig. 2 to the Curie-Weiss law, is shown as a function of composition in Fig. 3. A temperature independent term, $\chi_0$, is included in the fits and determined to a precision of $\pm 5 \times 10^{-5}$ emu/mol Ru. The substitution of Ti for Ru systematically shifts $\theta_{CW}$ towards positive values for $x \leq 0.4$, changing from $\theta_{CW} = -162$ K for $x=0$ to $\theta_{CW} = -37$ K for $x=0.4$. The composition dependence of the effective magnetic moment, $\mu_{eff}$, also derived from the data in Fig. 2, is shown in the inset of Fig. 3. The $\mu_{eff}$ per mole $CaRu_{1-x}Ti_xO_3$ formula unit decreases smoothly with increasing Ti content, as expected from the non-magnetic nature of $Ti^{4+}$. The $\mu_{eff}$ per mole Ru, however, stays constant across the series, at approximately 2.8 $\mu_B$/Ru, corresponding to the value expected for a low spin $Ru^{4+}$ (S=1) configuration. This result is in accord with the fact that $CaRu_{1-x}Ti_xO_3$ is an isovalent doping system. The Ru atom still has a valence of 4+ and it is still in an octahedral site. The only change is that the long-range three-dimensional Ru-O-Ru network in $CaRuO_3$ is disrupted by disorder, as the B-site is occupied both by conducting and magnetic $Ru^{4+}$ and by insulating and nonmagnetic $Ti^{4+}$.

The disorder introduced in B sites has a significant yet unexpected effect on the magnetic properties. Though the meaning of the Curie-Weiss temperature in a complex magnetic system such as this is not clear, it can be considered as an order parameter for characterizing the progression between classical ferromagnetic and non-ferromagnetic behavior. [18] There is a clear trend towards increasing ferromagnetism on Ti-doping, as can be seen by the more positive Curie-Weiss temperatures. This trend is further confirmed by the magnetization measurements. The main panel of Fig. 4 shows the magnetic hysteresis loops measured for a variety of $CaRu_{1-x}Ti_xO_3$ compositions at 5 K, and the lower left inset shows the data for x=0.02. It is remarkable that as little as 2% Ti substitution creates a hysteresis loop, which indicates that the magnetic ground state has a significant ferromagnetic component. There are no other sources of possible ferromagnetism, i.e., all possible impurities like $CaRuO_3$, $CaTiO_3$, $TiO_2$, $RuO_2$ or $Ca_3Ru_2O_7$, are not ferromagnets. This ferromagnetism is therefore an intrinsic property of the compound. The sizes of the loops increase until x equals 0.4, and then decrease as the samples approach the non-magnetic end member $CaTiO_3$. It is important to note that the change in cell dimension and distortion of $CaRuO_3$ on Ti doping (inset to Fig. 1) is opposite to the effect observed in the $Ca_{1-x}Sr_xRuO_3$ series. [19] Therefore the induced ferromagnetism cannot be explained by the arguments often employed to explain the difference between "paramagnetic" $CaRuO_3$ and ferromagnetic $SrRuO_3$. [10-12] We note that unlike in the present case, where ferromagnetism has been induced in $CaRuO_3$ by doping with non-magnetic Ti, no ferromagnetism was reported in the $CaRu_{1-x}Sn_xO_3$ series, [20] probably due to a very large disruption of the Ru-O sublattice by chemically and metrically dissimilar Sn. The substitution of Rh in $CaRuO_3$ has been reported, [21] but Rh is expected to change both the oxidation state of Ru and to have a magnetic contribution as well.

The magnetization was measured as a function of magnetic field (M-H curve) in the temperature range 5-55 K for each sample. All M-H plots show straight lines between 1 T and 2 T. At each temperature, the ferromagnetic contribution to the magnetization, $M_F$, was obtained by extrapolating the magnetization at higher field to zero field. The inset in the upper left panel of Fig. 4 shows $M_F$ as a function of temperature for x=0.2, 0.4, 0.6, and 0.8. For each sample, the sizes of the hysteresis loop decrease with increasing temperature, as reflected in the decreasing $M_F$. The temperature at which ferromagnetism turns on is, within our precision, independent of the Ti doping content, i.e., all loops are closed at 55 K, and a hysteresis loop can be observed only at lower temperature. Therefore the ferromagnetic $T_C$ is 55 K and independent of composition from x=0.2 to x=0.8. The magnitude of the largest $M_F$/Ru, ~ 530 emu/mol Ru at 5 K for x= 0.4, is considerably lower than that observed for $SrRuO_3$, which is in the range of ~ 4700-8900 emu/mol Ru. [22] These magnetizations are consistent with an itinerant electron picture of ferromagnetism in both materials. [14-16] Preliminary neutron diffraction experiments at 13 K on $CaRu_{0.5}Ti_{0.5}O_3$ showed that there are no additional magnetic neutron scattering peaks, indicating that these materials are not canted antiferromagnets. [23]

## IV. CONCLUSION

The magnetic ground state of $CaRuO_3$ is problematic. Once thought to be a conventional paramagnetic metal with antiferromagnetic spin interactions, it has recently been proposed to be a nearly ferromagnetic metal, for which a classical Curie-Weiss type interpretation of the magnetic susceptibility is not valid. In the current study, samples of $CaRu_{1-x}Ti_xO_3$ ($0 \leq x \leq 1$) were prepared and studies of their magnetic properties show that $CaRuO_3$ can be made to display ferromagnetic behavior by introducing very small amounts of disorder onto the Ru sites by partial substitution of Ru by non-magnetic, isovalent Ti. All Ti-doped samples show magnetic hysteresis loops below 55 K and the substitution systematically shifts the Curie-Weiss temperatures towards positive values. The disorder induced ferromagnetism cannot be explained in a conventional band-structure based picture, as the disorder would be expected to decrease the density of states at the Fermi level, favoring a non-magnetic ground state. [11-12] Rather the data imply that $CaRuO_3$ itself is poised at a critical point between ferromagnetism and paramagnetism at low temperatures, a balance tipped in favor of the former by small amounts of disorder. This strongly supports the proposal that the magnetic ground state of $CaRuO_3$ is that of a nearly ferromagnetic metal, not a classical antiferromagnet. Thus, conventional magnetic behavior in the family of perovskite-based ruthenates remains elusive.

### ACKNOWLEDGMENTS

This work was supported by the National Science Foundation grant No. DMR-9725979.

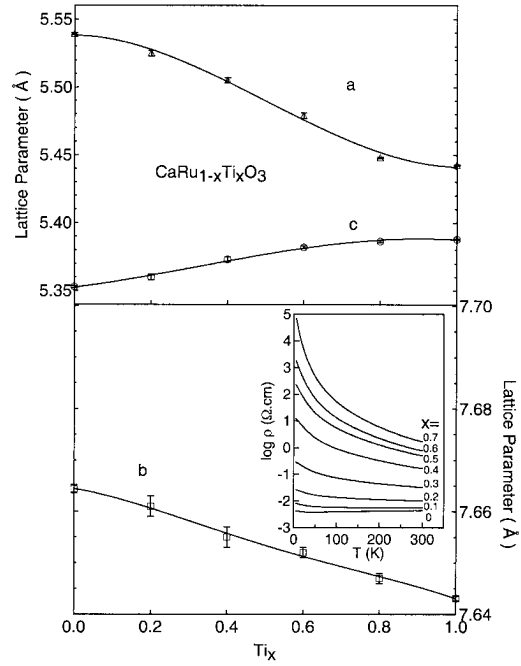

FIG. 1. Unit cell dimensions for orthorhombic $CaRu_{1-x}Ti_xO_3$ as a function of Ti doping content, solid lines are guides to the eye. Inset: Temperature dependence of the electrical resistivity of polycrystalline samples of $CaRu_{1-x}Ti_xRuO_3$.

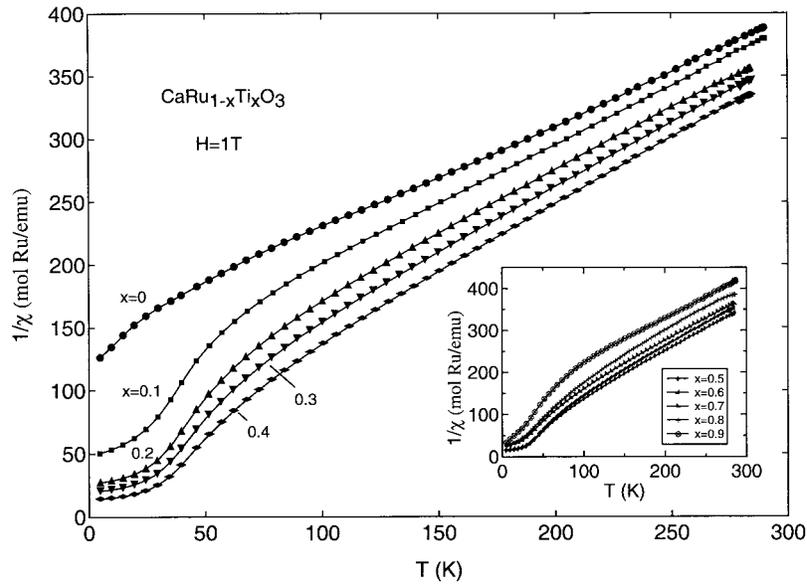

FIG. 2. Temperature dependence of the inverse magnetic susceptibility measured at 1 T on field cooling for $CaRu_{1-x}Ti_xRuO_3$ ($0 \leq x \leq 0.4$). Inset: same data for $0.4 \leq x \leq 0.9$.

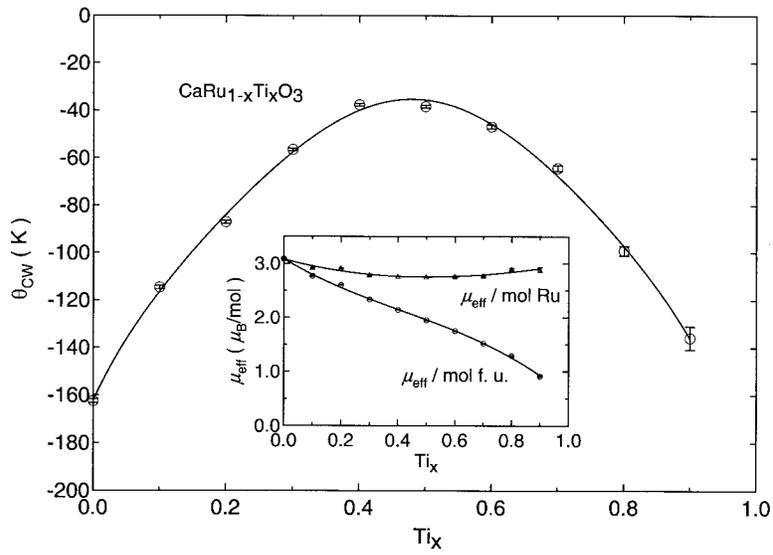

FIG. 3. Composition dependence of Curie-Weiss temperatures for $CaRu_{1-x}Ti_xRuO_3$. Inset: Composition dependence of the effective magnetic moment calculated in per mole formula unit (○) and in per mole Ru (Δ). Solid lines are guides to the eye.

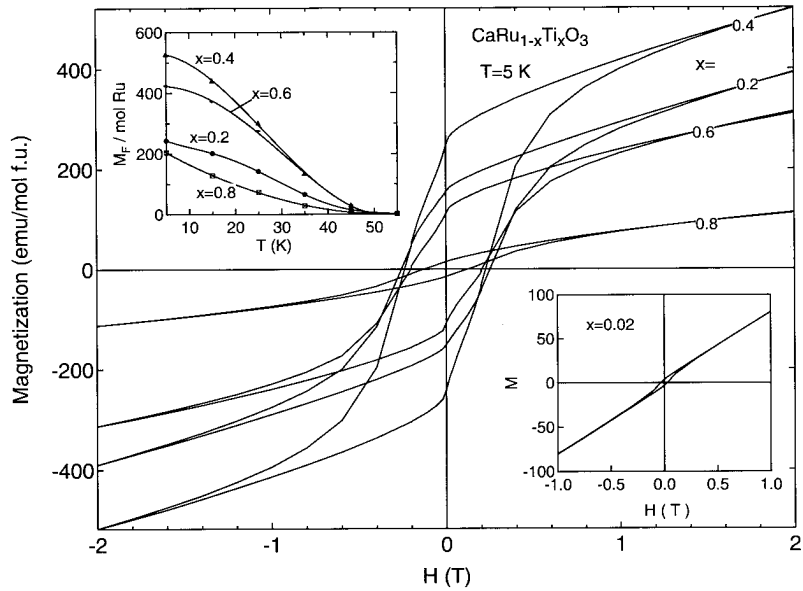

FIG. 4. Magnetic hysterisis loops for $CaRu_{1-x}Ti_xRuO_3$ at 5 K for x=0.2, 0.4, 0.6 and 0.8. Inset, upper left: Temperature dependence of the ferromagnetic contribution to magnetization for $CaRu_{1-x}Ti_xRuO_3$ with x=0.2, 0.4, 0.6 and 0.8. Inset, lower right: Magnetic hysterisis loop for $CaRu_{0.98}Ti_{0.02}RuO_3$ at 5 K.